# Effect of Spatial Heterogeneity on Near-Limit Propagation of a Stable Detonation

Jianling Li
Institute of Fluid Physics
China Academy of Engineering Physics, Mianyang, China

Xiaocheng Mi, Andrew J. Higgins
Department of Mechanical Engineering
McGill University, Montreal, Quebec, H3A 2K6 Canada

(Dated: 3 June 2014)

**Abstract**

The effect of introducing a spatial heterogeneity into an explosive medium is studied computationally by examining the detonation velocity near the limit to propagation in a thin explosive layer. The explosive system studied is an ideal gas with a single exothermic reaction governed by a pressure-dependent reaction rate ($p^n$) with a pressure exponent of $n = 3$. A pressure-dependent reaction rate, rather than the exponential dependence of reaction on temperature of Arrhenius kinetics, is used so that the detonation wave is stable in the homogeneous case and can be modelled with simple, analytical techniques, and thus the effect of introducing heterogeneity can be clearly identified. The two-dimensional slab of explosive is bounded by a layer of inert gas with the same thermodynamic properties as the explosive. The heterogeneity is introduced into the explosive via a large-amplitude, two-dimensional sinusoidal ripple in density in the initialization of the simulation, while maintaining a constant pressure. The computational simulations are initialized with a ZND solution for the ideal CJ detonation, and the detonation is allowed to propagate into the explosive layer. The simulations show that the detonation in the heterogeneous media exhibits a cellular-like structure of complex shock interactions. The detonation is able to propagate into a significantly thinner layer of explosive and can exhibit a greater velocity than the corresponding homogeneous case. A parametric study of varying the wavelength of the sinusoid shows the existence of an optimal size of heterogeneity at which the favorable effect is the greatest corresponding to a wavelength that is approximately 10 to 50 times the half reaction zone length of the ideal CJ detonation. As the wavelength is decreased to the size of the reaction zone length, the behavior reverts back to the homogeneous case.

**Keywords**

Detonation; heterogeneity; inhomogeneity; non-ideal detonation; pressure-dependent reaction

## 1. Introduction

The propagation of detonation waves in explosive media is dominated by the spatial heterogeneity of the media or the heterogeneous structure of the detonation wave itself. In polycrystalline solid explosives, the initiation and growth of reaction is controlled by localization of the shock energy at heterogeneous centers such as pores and explosive grain boundaries, commonly referred to as "hot spots" in the explosives literature. In homogeneous explosives, such as liquid explosives, while the media itself is uniform, the temperature sensitivity of Arrhenius governed kinetics results in the wave structure being dominated by spatial and temporal instabilities that are usually manifested as a cellular structure of the wave. Despite the importance of these features, they are typically neglected in models of detonation propagation in condensed phase explosives or are embedded in semi-empirical models for the reaction rate that are used in computational simulations that assume a spatially uniform media [1]. In gaseous explosives with Arrhenius kinetics, unsupported two dimensional detonations (i.e., detonation waves that are not overdriven by a piston) can be shown to be unstable to transverse perturbations for all values of activation energy [2,3]. This instability is manifested as a cellular structure in detonation propagation, and the cellular structure is well acknowledged as the basis of successful empirical correlations for detonation initiation and propagation [4,5]. Computational simulation of cellular structure is extremely challenging due to the numerical resolution necessary to resolve the fine scales of the cells and the extreme sensitivity of activated reactions, as exemplified by the recent study of Mazaheri et al.[6] In their study, two-dimensional computational simulations of detonations with activation energies representative of real gaseous mixtures exhibited increasingly finer and more complex structures (shear layers, vortical structures, shock interactions, etc.) as the resolution of the computational grid was increased to as many as 1000 computational cells per half reaction zone length, and these structures were seen to play a significant role in the burning mechanism of the detonation wave. Thus, fully resolved simulations of gaseous detonations with detailed chemistry are likely to remain computationally challenging for the foreseeable future.

In this study, an initially stable detonation wave governed by a pressure dependent reaction rate is used to examine the effect of spatial heterogeneity of the detonable media. The use of a pressure-dependent reaction rate ($r \sim p^n$) avoids the stability and computational grid resolution issues associated with using an Arrhenius-based reaction rate. However, critical phenomena (i.e., detonation failure) is still observed if the detonation experiences losses, such as divergent flow due to yielding confinement, provided that the pressure exponent is sufficiently large ($n \geq 2$) [7]. The existence of critical phenomenon, such as a critical charge thickness required for propagation, allows the influence of heterogeneity upon detonation sensitivity to be studied. The variation of density within the explosive was identified as a measure of heterogeneity in an early computational study by Mader [8]. The spatial heterogeneity is introduced in the present study via a sinusoidal ripple in density that extends through the detonable media, maintaining uniform pressure. Similar sinusoidal ripples have been introduced in simulations to examine the influence upon shock ignition by Stewart [9]. His results showed that, due to the presence of the heterogeneities, the induction time of thermal explosion following a shock is reduced by about one half. Perhaps the closest study to the current one is that of Morano and Shepherd [10], who performed one-dimensional computations of detonation in an idealized system with a pressure-dependent reaction rate that was meant to represent a solid explosive. A spatial, sinusoidal variation in the reaction rate constant was introduced and the effect of the variation of amplitude and wavelength was studied. For large wavelengths and amplitudes, a small, negative deviation in detonation velocity away from the ideal CJ velocity was observed, but this deficit did not exceed 2%.

The present study extends this approach to two dimensional variations of explosive properties in layers with yielding confinement, so that critical, near-limit behavior may be examined. The variation is applied to the density field (as opposed to the reaction constant as in the Morano and Shepherd study) while the pressure is maintained as constant, resulting in the temperature field having an initial variation as well. The resulting media has a heterogeneous acoustic impedance that will result in local shock diffraction and focusing, leading to shock interactions such as Mach reflections. These nonlinear interactions can feedback to the energy release required to support the detonation via the pressure-dependent reaction rate. Thus, the use of an imposed, controlled variation in properties can be a prototype system to explore the influence of both heterogeneities (such as encountered in polycrystalline explosives) and intrinsic instabilities (such as occurs in liquid and gaseous homogeneous explosives). The full spectrum of wavelength of these heterogeneities, from wavelengths much larger than the reaction zone length to wavelengths smaller than the reaction zone length, is examined in this study. The effect on detonation propagation will be quantified by examining the critical thickness of a 2-D slab of explosive bounded by an inert media and the velocity deficit for detonation propagation near that critical thickness, conceptually similar to experiments with detonations in a layer or column of explosive gas bounded by a layer of inert gas performed by Sommers and Morrison [11] and Vasil'ev and Zak [12] .

## 2. Problem Description and Numerical Technique

### 2.1 Problem Description

The system studied is a calorically perfect gas with a single exothermic reaction, with the reaction rate given by

$$\frac{dZ}{dt} = k(1-Z)\left(\frac{p}{p_{CJ}}\right)^n \quad (1)$$

Where $Z$ is the reaction progress variable ($Z = 0$ for unreacted, $Z = 1$ for reacted). The pressure exponent of $n = 3$ was selected in this study since it is representative of reaction rate models used for solid explosives, results in a critical velocity, and is likely to accentuate enhanced reaction rates that occur in regions of high pressure resulting from shock interactions. The thermodynamic properties of the explosive are as follows, $\Delta q/RT_1 = 24$ and $\gamma = 1.333$. We emphasize that the use of a pressure-dependent reaction rate means that these simulations are not intended to represent any real gaseous mixture. In many ways, this study is more relevant to condensed-phase explosives, with the ideal gas law being used for convenience and simplicity.

The detonation is initialized in a uniform region with a ZND detonation wave profile for the ideal CJ detonation (i.e., detonation without losses) and allowed to propagate into a two-dimensional layer bounded on one side by a layer of the same gas at the same initial temperature, density, and pressure, but made non-reactive by setting $Z = 1$ in that region (see Fig. 1). The other boundary invoked mirror boundary conditions to reflect the symmetry of the problem. Simulations were also performed with the explosive layer confined on both sides by a solid wall boundary condition (no inert layer) in order to examine the effect of the heterogeneity upon propagation without losses.

The heterogeneity was introduced via a variation in the initial density of the explosive medium as given by

$$\rho = \rho_1 \left\{ 1 + \sigma \left[ \sin(\frac{2\pi}{\lambda}(x - x_1)) \cos(\frac{2\pi}{\lambda} y) \right] \right\} \qquad (2)$$

where $x$ is the direction of detonation propagation, $x_1$ denotes the location of the transition from the initiation region to the region bounded by inert gas, and $y$ is the direction perpendicular to propagation. The amplitude of the variation $\sigma$ was fixed at 0.5 for this study, and the wavelength of the sinusoid, $\lambda$, was varied over two orders of magnitude in different simulations. The particular combination of sine and cosine used in Eq. (1) ensured that the uniform density field in the initiation region continuously matched the heterogeneous region and that the variation in the *y*-direction was symmetric about the *x*-axis. A subtle detail of this variation is that the temperature must vary with the reciprocal of the density if the pressure is held constant, which results in a non-sinusoidal variation in temperature. However, the total internal energy of the gas divided by its total heat capacity resulted in the same temperature as the homogeneous case, and the total mixture chemical energy content remained the same as well. This system can be thought of as a gaseous explosive wherein pockets of cold, dense gas mixture alternate in a chessboard pattern with pockets of warm, low-density gas, while the total pressure and energetic content of the gas remains constant in comparison to the equivalent homogeneous gas mixture.

**2.2 Numerical Method**

The governing unsteady, two-dimensional inviscid Euler equations with the same pressure-dependent reaction rate source term described above were solved on a uniform, Cartesian computational grid. The computations were performed in the lab-fixed reference frame, with the computational domain reinitialized at finite intervals to always contain the leading shock front and downstream limiting characteristic within the domain. The finite-volume algorithm and a third-order TVD Runge-Kutta method [13] was used for spatial and temporal discretization. The method reverts to first-order in the presence of discontinuities such as shock waves. The AUSM+ scheme [14], based on the fact of convection and acoustic waves as two physically distinct processes, was used to deal with the inviscid flux as a sum of the convective and pressure terms. The source term was dealt with by using a second-order accurate Strang operator splitting method [15] and fully implicit method. The boundary condition along the *x*-axis was a mirror boundary condition (axis of symmetry), so that only the upper half of the layer is simulated in the case of a two-dimensional slab. The upper boundary of the computational domain (above the inert layer) was a supersonic outflow condition to ensure that no reflected waves return into the computational domain. A study of the variation of the thickness of the inert layer confirmed that it was sufficiently thick to have no influence on the results reported below.

## 3. Results

### 3.1 Homogeneous Case with Yielding Confinement

The behavior of detonation waves in this system for the case of an initially homogeneous explosive (i.e., no initial sinusoidal variation in properties) has been thoroughly characterized in the study of Li et al. [16]. Computational simulations in this case show detonation waves that are stable and, after an initial transient decay, propagate at constant velocity, with the shock front being smooth and having a well characterized curvature, as shown in Fig. 2(a). The steady-state propagation velocity measured from simulations with different layer thicknesses $t$ (normalized by the half reaction zone thickness of the ideal CJ detonation, $L_{1/2}$) in this case is reported in Fig. 3 and compared to a simple, analytic model based on front curvature. The reciprocal of the charge thickness is use for the $x$-axis so that extrapolation to the $y$-axis yields the ideal CJ detonation velocity without losses, and a factor of two is included ($L_{1/2}/2t$) so that results with cylindrical charges plotted using the charge diameter would lie on the same curve, following the convention of plotting diameter-velocity relations in the condensed-phase explosives literature. The velocities reported here were shown to be grid independent, requiring only three computational cells per half reaction zone thickness to obtain a numerically converged result for velocity [16].

Detonation waves of this type can be treated using a simple curvature-based model. The analytic model uses the detonation velocity-front curvature relation obtained by integrating through the one-dimensional reaction zone with an area divergence term (following Wood and Kirkwood [17, 18]). The shock curvature is assumed to vary continuously across the shock front using a geometric construction first proposed by Eyring et al. [19] and is then matched via shock polar analysis to the shock angle at the boundary with the inert confinement. The study by Li et al. [16] can be consulted for further details of this computation and modelling.

In Fig. 2, the velocity is seen to decrease as the explosive layer is made thinner until a critical velocity of approximately 80% of the ideal detonation velocity is encountered, at which point the detonation in the simulations promptly fails and the analytic solution exhibits a turning point. The fact that a simple, curvature based model can successfully reproduce the detonation velocity-thickness relationship and predict the critical thickness is the reason why this particular system was selected for further study on the effect of introducing heterogeneity. The region of the plot in Fig. 3 near the critical thickness ($L_{1/2}/2t = 0.002555$, or a layer thickness $t = 195.7 L_{1/2}$) was selected for the study of the effect of introducing heterogeneity in Section 3.3 below.

**3.2 Heterogeneous Case with Rigid Confinement**

The effect of an imposed heterogeneity in the propagation of a detonation with rigid confinement is examined by two-dimensional simulations with mirror boundary condition on both upper and lower sides of the computational domain and one-dimensional simulations with a sinusoidal variation of initial density only in the *x*-direction. In Fig. 4, it can be seen that, for a wide range of $\lambda$ spanning nearly two orders of magnitude, the resulted detonation velocities in one and two-dimensional simulations only deviate from the ideal CJ velocity by less than 1.2%. This result is comparable to that obtained by Morano and Shepherd's one-dimensional simulation in which heterogeneity had been introduced via a sinusoidal variation of reaction rate [10]. The salient result of these simulations is that the introduction of a sinusoidal variation in density has not significantly affected the energetics or ideal detonation velocity of the explosive medium in the case of rigid confinement.

## 3.3 Heterogeneous Case with Yielding Confinement

To examine the effect of an imposed heterogeneity in the propagation of a detonation in a near limit case with yielding confinement, a layer with thickness $t = 195.7 L_{1/2}$ was selected (see Fig. 2). The detonation was initialized with an ideal CJ detonation (the same as the homogeneous cases in Section 3.1) and then allowed to propagate into the region with the spatial sinusoid. The different wave structures observed are shown in Fig. 2 (b)-(f) for selected values of $\lambda$. The wave structure takes on the appearance of a cellular-like detonation front, but this structure is a result of the imposed sinusoid. In Fig. 2, it can be seen that, while the media is initially uniform in pressure, upon passage of the shock front, significant pressure amplification can be associated with the spatial variation in material density and impedance. Due to the large pulsations of the shock front, particularly in the case of large $\lambda$, it was necessary to track the wave for a considerable distance (a distance of 2000 $L_{1/2}$, corresponding to approximately 10 thicknesses of the charge) in order to obtain an effective average velocity. The average wave velocity obtained from the simulations with various values of $\lambda$ is shown in Fig. 5 (for both linear and log representations of the *x*-axis), with the wave velocity being normalized by the steady state velocity of the corresponding homogenous case with the same computational resolution. We emphasize that these velocities are not the ideal, Chapman-Jouget detonation velocities of an effectively infinite diameter charge, but rather the non-ideal velocities observed near the limit to propagation in a charge with significant lateral losses due to the yielding confinement.

The influence of the computational grid resolution was studied by performing all simulations at four different resolutions: 3, 6, 7.5, and 9.5 computational cells per half reaction zone thickness of the ideal detonation. The results are compared in Fig. 5 using different symbols, with curves added to assist the eye in identifying trends for each set of simulations performed at a given resolution. The overall results are similar, at different resolutions, with values of $\lambda$ ranging from 10 $L_{1/2}$ to 50 $L_{1/2}$ resulting in a significantly greater propagation velocity than the corresponding homogeneous case. As the computational grid resolution was increased, the effect becomes more significant, with the wave velocity increasing and the range over which an augmented wave velocity is observed extends to lower values of $\lambda$. For values of $\lambda$ approaching $\lambda = L_{1/2}$, simulations performed at all resolutions appear to converge back to the homogeneous result, evidence that for a very small wavelength of heterogeneity, the media behaves as if homogenous. The results of the two highest resolution series of simulations (7.5 and 9.5 computational cells per half reaction zone) agree to within the accuracy of the measured velocity, so these results seem to have achieved numerical convergence. As the wave length is increased (greater than 60 $L_{1/2}$), the scale of the inhomogeneity becomes comparable to the size of the explosive layer. The computations were not extended to this limit, as the detonation would fail as it encountered a large pocket of lower density explosive.

In order to investigate the effect of heterogeneity upon the ability of a detonation to propagate in a thinner layer, a wavelength of $\lambda = 40\ L_{1/2}$ was selected for the imposed heterogeneity, and simulations conducted with different thicknesses of the detonable layer with a computation resolution of 3 computational cells per half reaction zone thickness. The results are shown in Fig. 6 in comparison to the results obtained in the homogeneous case (cf. Fig. 2). The detonation wave is able to propagate into a much thinner layer of explosive (143.5 $L_{1/2}$) and at a greater velocity in comparison to the homogenous case due to the presence of the imposed heterogeneity.

**4. Discussion and Conclusion**

The presence of a sinusoidal variation in the density of an explosive media is seen to have a significant influence on the propagation of a detonation wave, enabling the wave to propagate at near limit conditions at greater velocities and in thinner layers than in comparison to the corresponding homogeneous case. The optimal wavelength of heterogeneity to promote successful propagation of detonation appears to be in the range of 10 to 50 times the half reaction zone length. While it is not presently known if this result is universal or specific to the particular range of parameters and numerical resolution used in this study, it is intriguing that this is the same size as the spatial instability observed in gaseous explosives in the form of cellular structure when the reaction rate is controlled by Arrhenius kinetics. Specifically, the transverse waves that define detonation cells in both experiments and simulations using activated kinetics are usually observed to have a transverse spacing of approximately 30 to 70 times the induction zone length as computed in a one-dimensional ZND-type model [4, 5].

In the present study, as the wavelength of the imposed inhomogeneity is made smaller (i.e., wavelength approaches the thickness of the reaction zone, $\lambda \approx L_{1/2}$), the dynamics of the wave appears to revert back to homogeneous behavior. Simulations $\lambda < L_{1/2}$ are nearly indistinguishable from the homogeneous case shown in Fig. 2(a) and exhibit the same wave speed as the homogeneous case to four significant digits. The scale of the heterogeneity has apparently become so fine in this limit so as to "homogenize" the explosive medium. In the other limit, as the wavelength of the sinusoid becomes significantly larger than the optimal value and approaches the thickness of the explosive layer itself, the detonation wave must propagate through shallow gradients in properties and encounters large regions of low density that appear to be harmful to the ability of the wave to sustain itself, and the wave is no longer assisted by the spatial inhomogeneity. In between these two limits, the presence of heterogeneities has a sensitizing effect on the detonation that assists its propagation. These two asymptotic behaviors are hypothesized to be the explanation for the existence of an optimal wavelength of heterogeneity observed in Fig. 5.

For a detonable media that has heterogeneity near the optimal wavelength, the detonation wave is able to propagate in a significantly thinner layer than the equivalent homogeneous system, as seen in Fig. 6. For the system studied here, the wave was able to sustain propagation in layers as thin as $143.5L_{1/2}$, in comparison to a critical thickness of 195.7 $L_{1/2}$ for the equivalent homogeneous system. The ability to propagate in significantly thinner layers without experiencing an additional deficit in velocity is strongly reminiscent of behavior observed in experiments with heterogeneous explosives featuring large scale heterogeneities by Petel et al. [20, 21].

Unlike simulations using Arrhenius kinetics, the use of a pressure-based reaction model should allow these simulations to be numerically converged with modest grid resolution requirements, and thus has the potential to circumvent the grid resolution issue that has frustrated computational simulations of detonations with Arrhenius kinetics for the last three decades. Extension of this study to three dimensions with spatially randomized perturbations in properties (thermodynamic, kinetic, or energetic) would be of interest, as it may give rise to percolating-like behavior wherein the detonation is able to exploit localized, random variations in properties to propagate in even thinner layers. The fact that randomized heat release can give rise to novel modes of flame propagation has been recently recognized as a new branch of combustion, "discrete combustion" [18-20]. The approach outlined in this paper may be a prototypical system to study the dynamics of detonations with spatially discrete, randomized energy release, such as encountered in non-ideal detonations in heterogeneous media as well as detonations in gases with highly irregular structure.

**Acknowledgments**

This work was supported by the National Natural Science Foundation of China (No. 51206150), National Key Laboratory for Shock Wave and Detonation Physics Research Foundation (No. 9140C6704010704) and the State Scholarship from China Scholarship Council.

**References**

1. E. L. Lee, C. M. Tarver, Phys. Fluids 23 (1980) 2362.

2. M. Short, D.S. Stewart. J. Fluid Mech. 368(1998) 229-262.

3. H.D. Ng, F. Zhang. Detonation instability, in: *Detonation Dynamics*, Shock Waves Science and Technology Library, volume 6, Springer, 2012, pp. 107-212.

4. J.H.S. Lee, Ann. Rev. Fluid Mech. 16 (1984) 311-36.

5. J.H.S. Lee, The Detonation Phenomenon, Cambridge, 2008.

6. K. Mazaheri, Y. Mahmoudi, M.I. Radulescu, Comb. Flame 159:6 (2012) 2138-2154.

7. M. Cowperthwaite, Phys. Fluids 6 (1994) 1357-1378

8. C. L. Mader, *Numerical Modelling of Detonations*, University of California Press, 1979

9. D. S. Stewart, Comb. Sci. Tech. 48:5-6 (1986) 309-330.

10. E. O. Morano, J. E. Shepherd, Effect of reaction rate periodicity on detonation propagation, in: *Shock Compression of Condensed Matter—2001*, AIP Conf. Proc. 620, 2002, pp. 446-449.

11. W.P. Sommers, R.B. Morrison, Phys. Fluids, 5 (1962) 241-248.

12. A.A. Vasil'ev, D.V. Zak, Comb. Expl. Shock, 22 (1986) 463-468.

13. C.W. Shu, S. Osher, J. Comput. Phys. 77 (1988) 429-471.

14. M.S. Liou, C.J. Steffen, J. Comp. Phys., 107 (1993) 23-39.

15. G. Strang, SIAM J. Numer. Anal. 5 (1968) 506-517.

16. J. Li, X. Mi, A.J. Higgins, Phys. Fluids, submitted 2014. http://arxiv.org/abs/1405.7753

17. W.W. Wood, J.G. Kirkwood, J. Chem. Phys. 22 (1954) 1920-1924.

18. Higgins, A.J., Steady one-dimensional detonations, in: *Detonation Dynamics*, Springer, 2012, pp. 33-105.

19. H. Eyring, R.E. Powell, G.H. Duffy, R.B. Parlin, Chem. Rev. 45 (1949) 69-181.

20. O.E. Petel, D. Mack, A.J. Higgins, R. Turcotte, and S.K. Chan, J. Loss Prevent. Proc. 20(4) (2007) 578-583.

21. O.E. Petel, D. Mack, A.J. Higgins, R. Turcotte, R., and S.K. Chan, Comparison of the detonation failure mechanism in homogeneous and heterogeneous explosives, in: *Proceedings of the 13th International Detonation Symposium*, 2006, pp. 2-12.

22. A.S. Mukasyan, A.S. Rogachev, Prog. Energy Comb. Sci., 34(3) (2008) 377-416.

23. S. Goroshin, F.D. Tang, A.J. Higgins, Phys. Rev. E, 84(2) (2011) 027301.

24. F.D. Tang, A.J. Higgins, S. Goroshin, S., Phys. Rev. E, 85(3) (2012) 036311.

**Figure Captions**

Figure 1       Schematic representation of the problem.

Figure 2       Structure of detonation wave for (a) homogeneous case, (b) heterogeneous case with wavelength of sinusoid $\lambda = 1\ L_{1/2}$, (c) $\lambda = 5\ L_{1/2}$, (d) $\lambda = 10\ L_{1/2}$, (e) $\lambda = 45\ L_{1/2}$, and (f) $\lambda = 80\ L_{1/2}$. The two-dimensional simulations are reflected about the horizontal axis, with shading representing density above the centerline and pressure below the centerline.

Figure 3       Detonation velocity (normalized by ideal CJ velocity) of a homogeneous explosive as a function of the thickness of a two dimensional layer bounded by inert gas. The symbols correspond to computational simulations and the curve is an analytic prediction based upon the Wood and Kirkwood model of the reaction zone to relate normal velocity to front curvature and matching the curvature to the shock interaction with the confinement [16].

Figure 4       Detonation velocities (normalized by ideal CJ velocity) from one-dimensional and two-dimensional simulations with rigid confinement (i.e., no lateral expansion) and a sinusoidal variation in density with wavelength $\lambda$.

Figure 5       Velocity of detonation propagation in simulations with sinusoidal variations in density, normalized by the detonation velocity of the corresponding homogenous case, as a function of the wavelength of the sinusoid $\lambda$, for (a) linear *x*-axis and (b) log *x*-axis. Simulations were performed at four different computational resolutions, with curves to assist in identifying trends.

Figure 6       Detonation velocity (normalized by ideal CJ velocity) of a heterogeneous explosive ($\lambda = 40 L_{1/2}$) as a function of the thickness of a two-dimensional layer bounded by inert gas, in comparison to results with a homogeneous explosive (cf. Fig. 2).

**Supplemental Material**

**S1** Animation_homogneous.avi: Animation of density field for simulation of detonation in homogeneous explosive media, as visualized from lab-fixed reference frame.

**S2** Animation_lambda=10.avi Animation of density field for simulation of detonation in explosive media with sinusoidal variation in density with wavelength $\lambda = 10\ L_{1/2}$, as visualized from lab-fixed reference frame.

**S3** Animation_lambda=45.avi Animation of density field for simulation of detonation in explosive media with sinusoidal variation in density with wavelength $\lambda = 45\ L_{1/2}$, as visualized from lab-fixed reference frame.

**S4** Animation_lambda=80.avi Animation of density field for simulation of detonation in explosive media with sinusoidal variation in density with wavelength $\lambda = 80\ L_{1/2}$, as visualized from lab-fixed reference frame.

Fig. 1

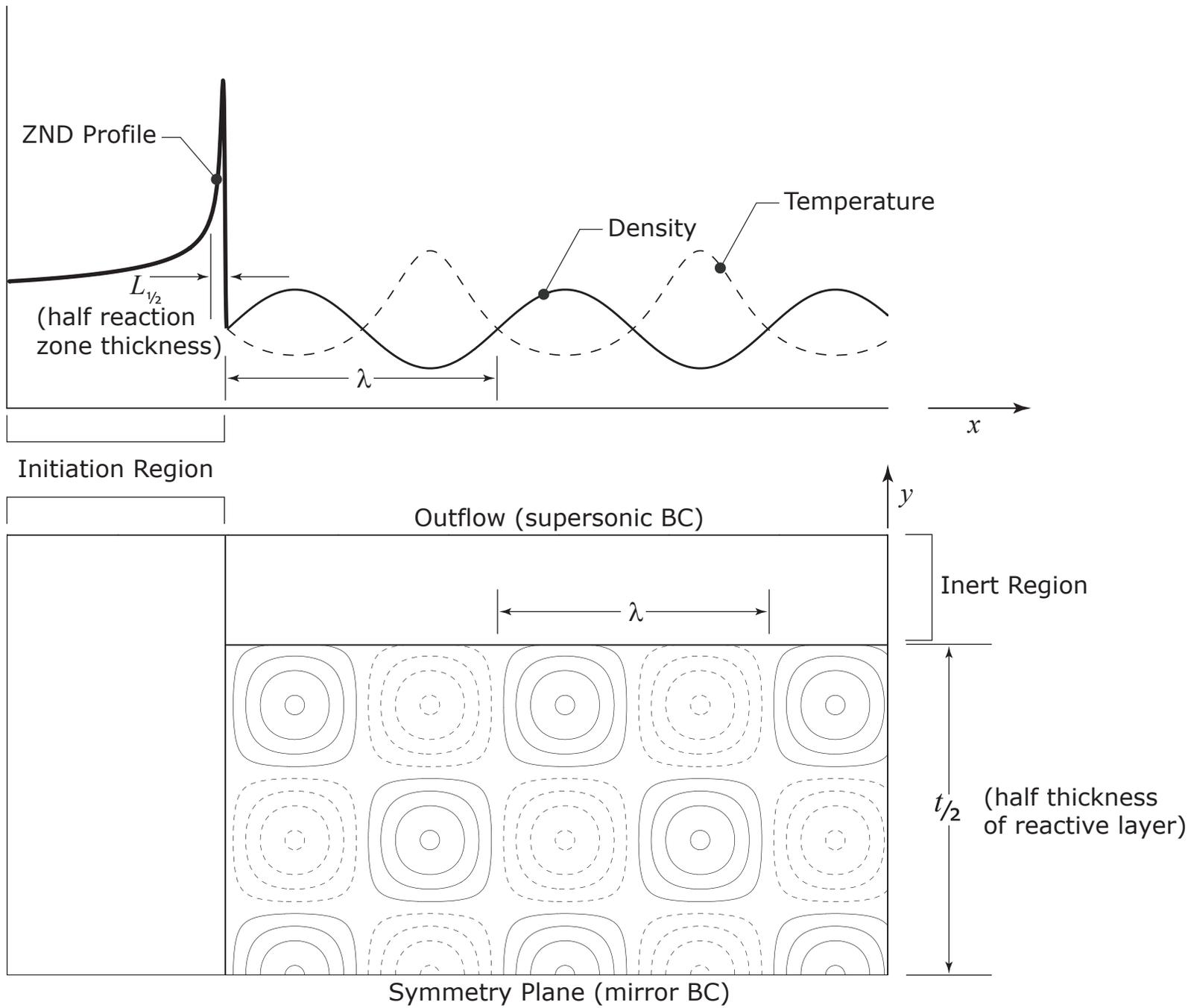

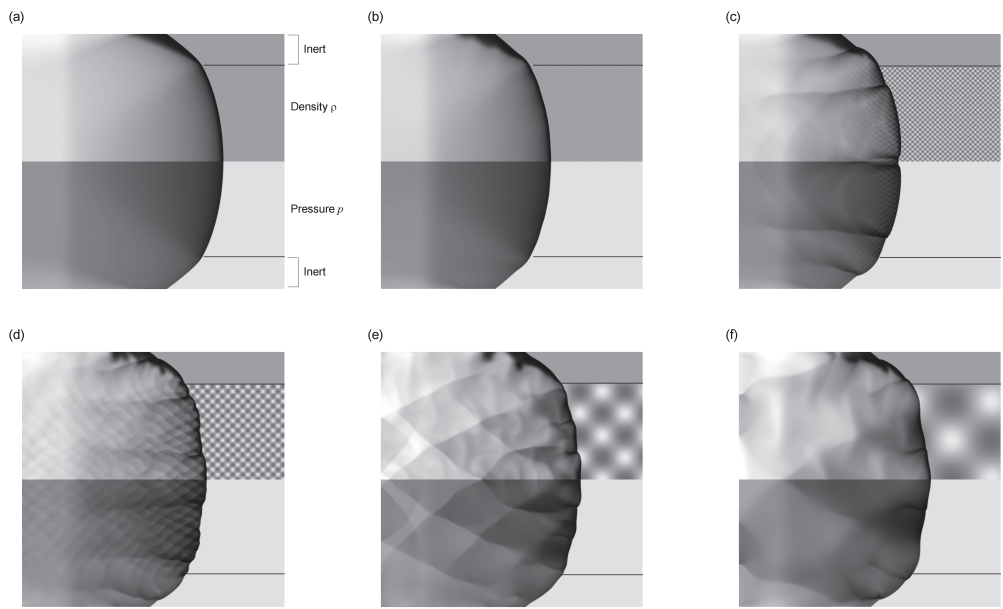

Fig. 2

Fig. 3

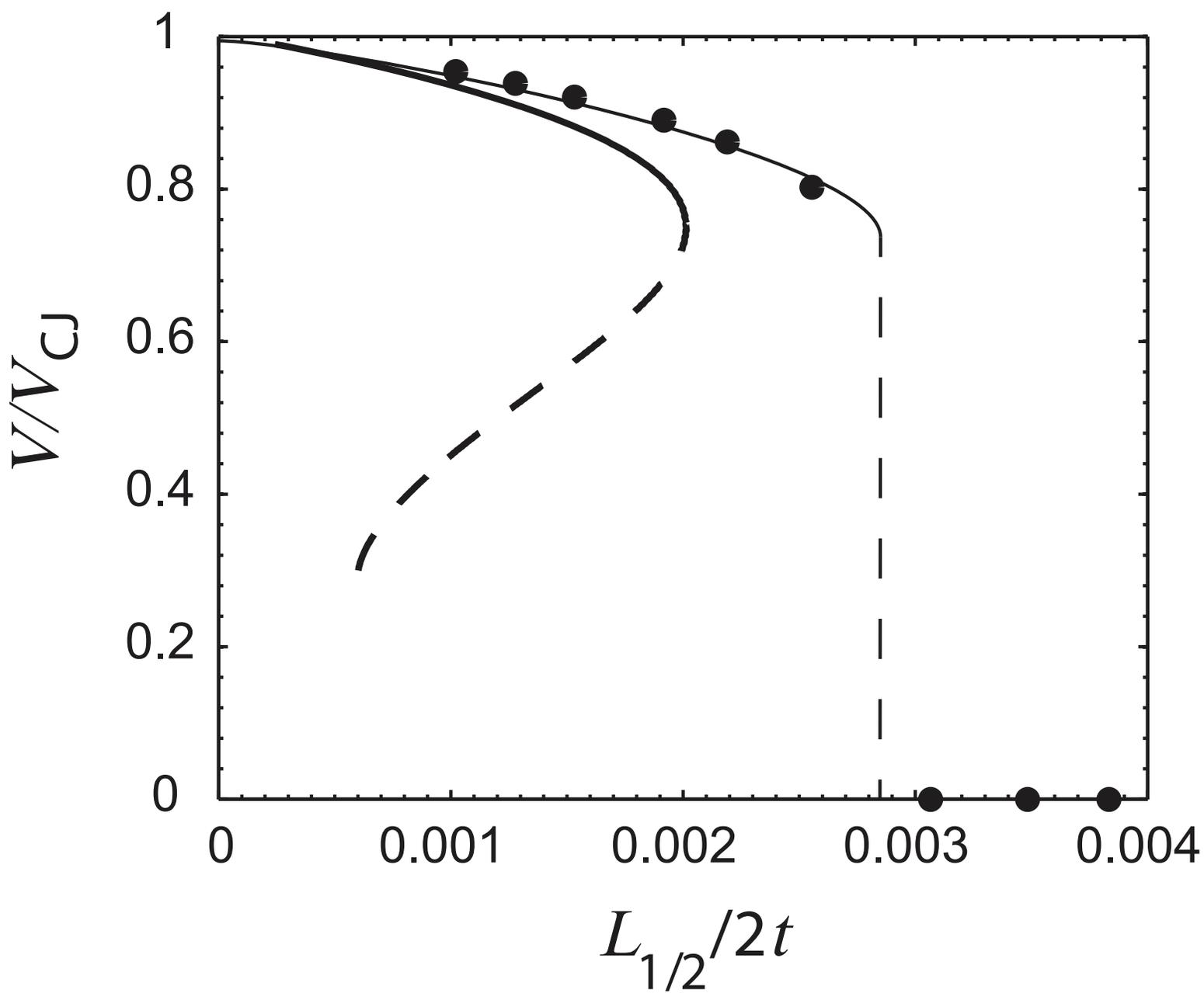

Fig. 4

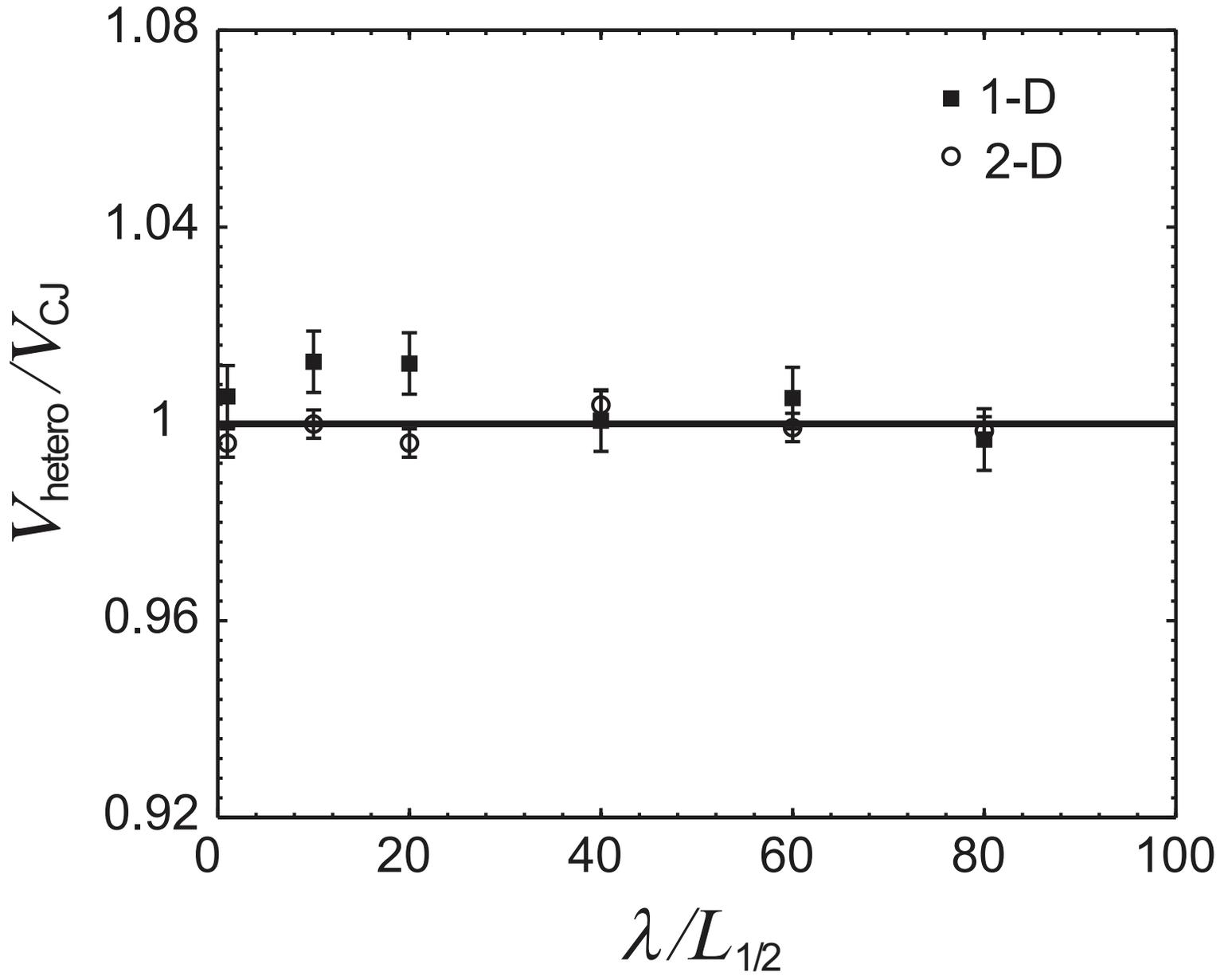

Fig. 5

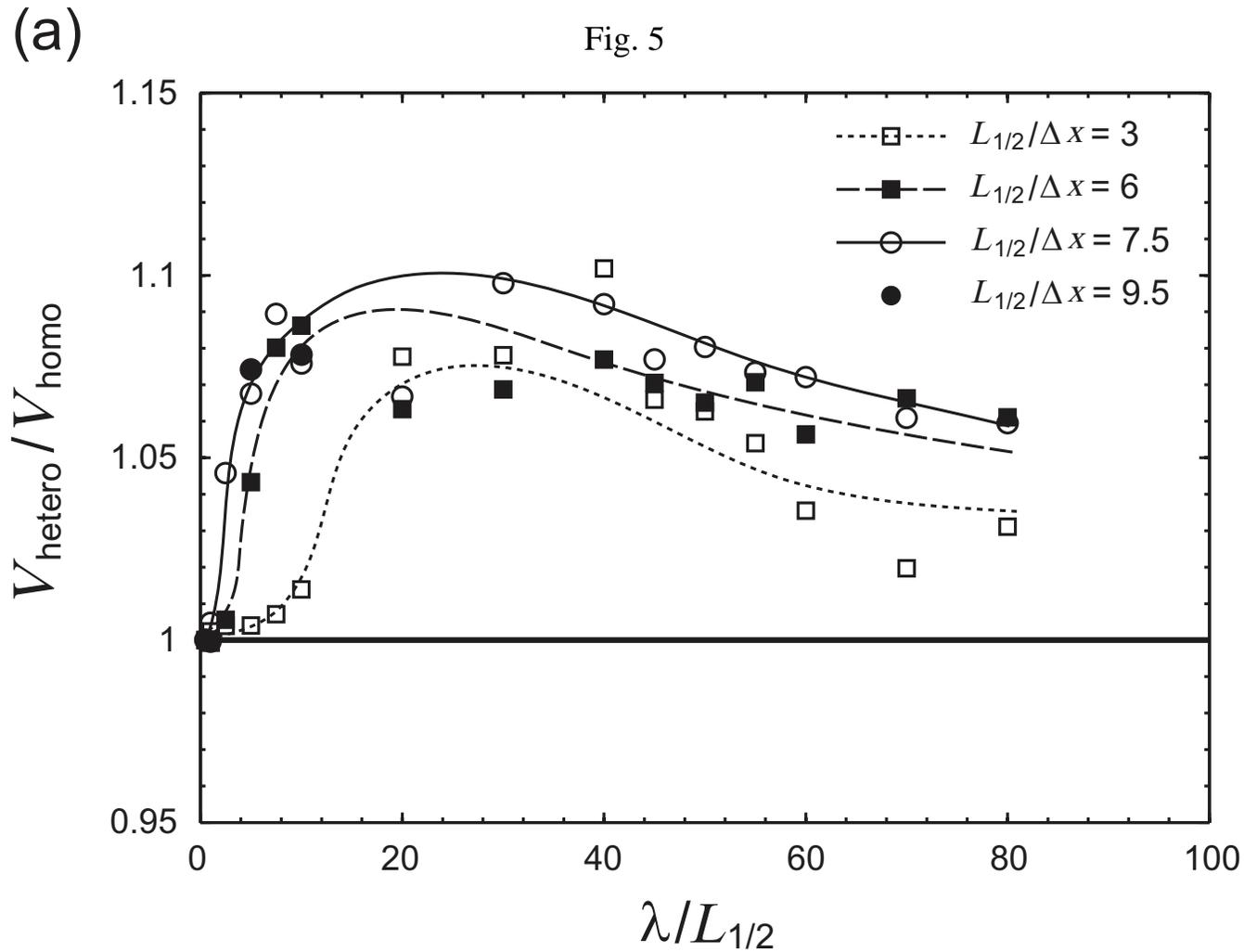

(a)

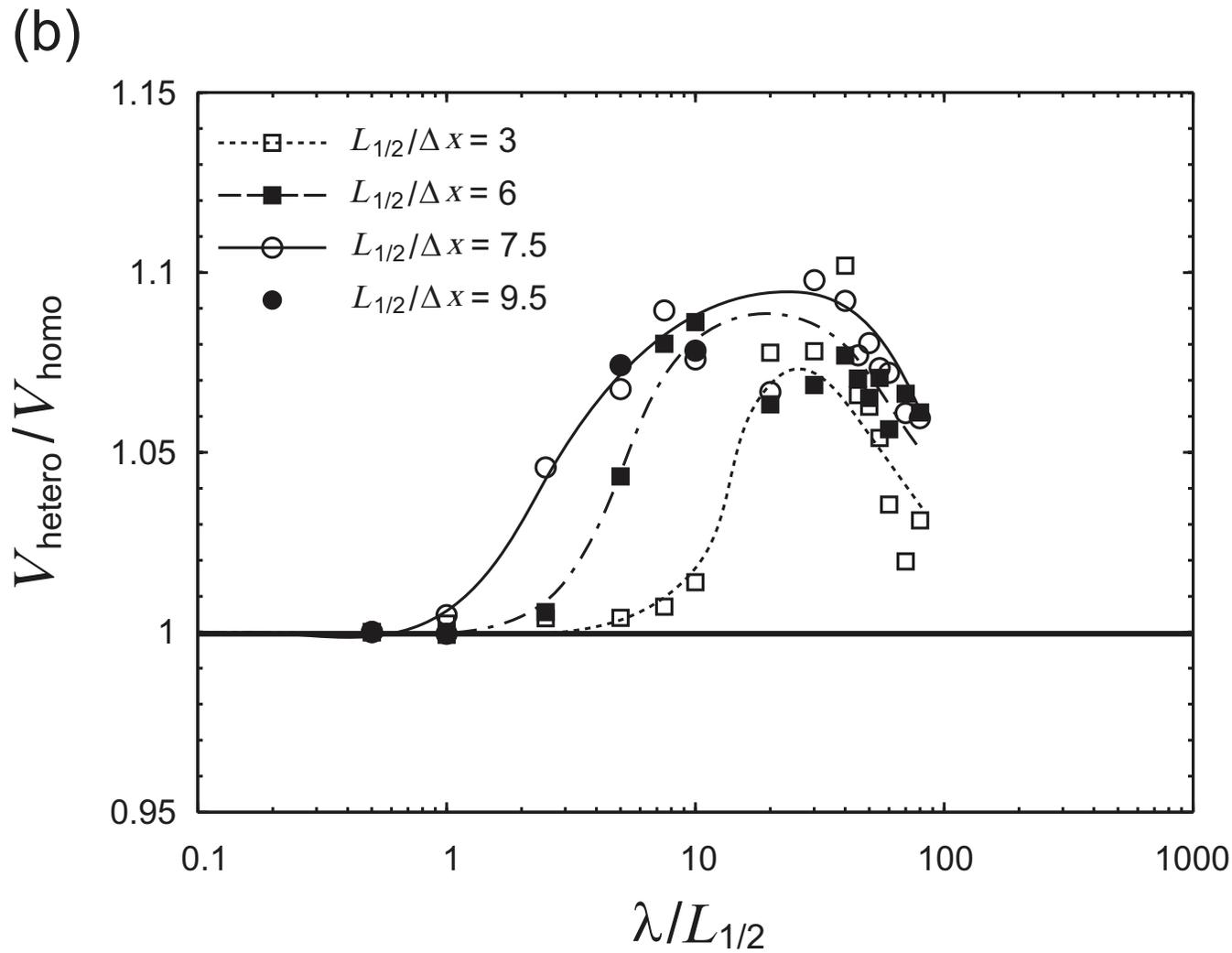

(b)

Fig. 6

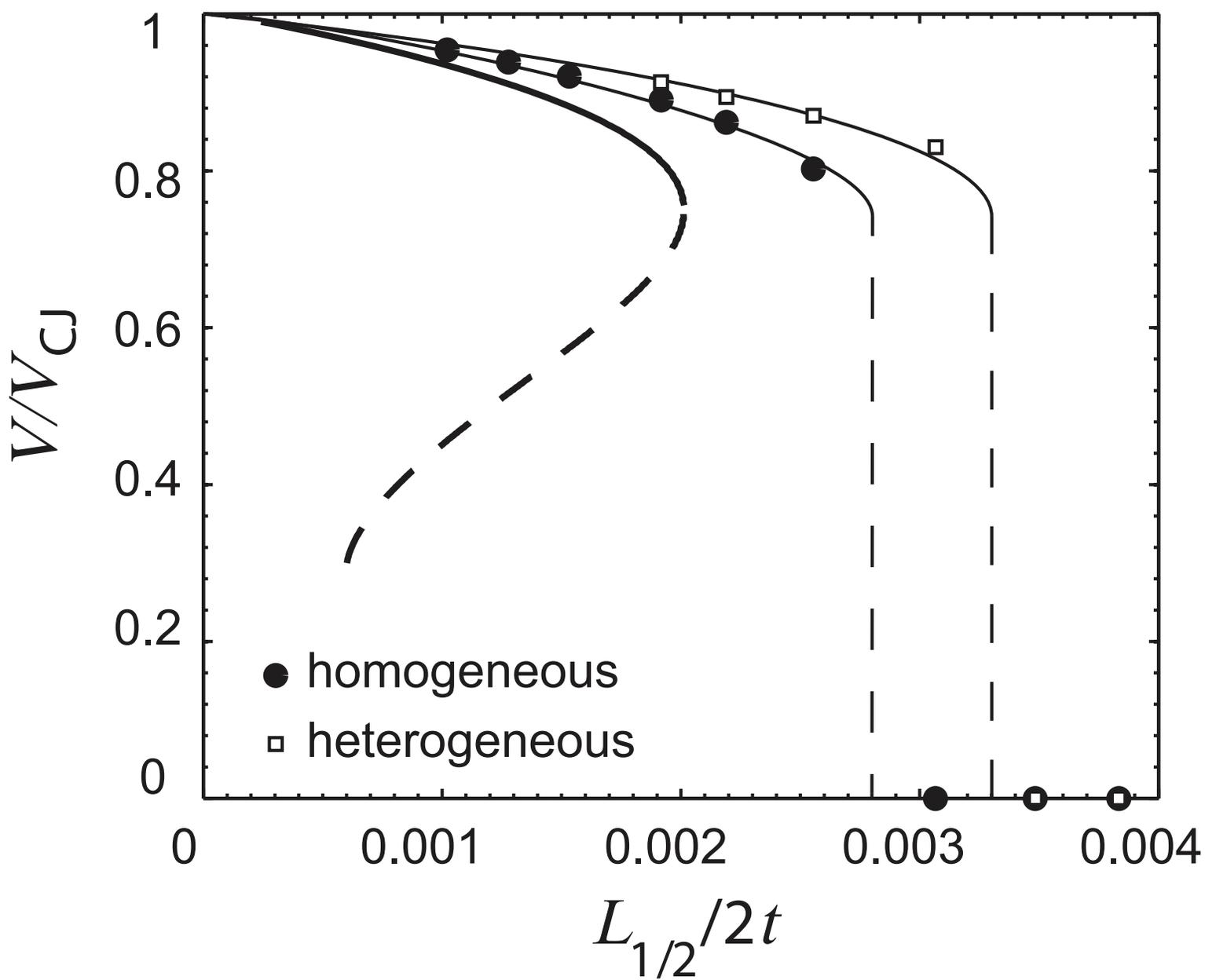